\documentclass{Interspeech2024}
\usepackage{array}
\usepackage{booktabs}
\usepackage{multirow}
\newcolumntype{C}[1]{>{\centering\arraybackslash}p{#1}} 
\newcolumntype{R}[1]{>{\raggedright\arraybackslash}p{#1}} 
\usepackage{subcaption}

\interspeechcameraready

\title{
Learn and Don't Forget: Adding a New Language to ASR Foundation Models}

\name[affiliation={}]{Mengjie}{Qian}
\name[affiliation={}]{Siyuan}{Tang}
\name[affiliation={}]{Rao}{Ma}
\name[affiliation={}]{Kate M.}{Knill}
\name[affiliation={}]{Mark J. F.}{Gales}

\address{
 ALTA Institute, Machine Intelligence Lab, Department of Engineering, Cambridge University, UK}
\email{\{mq227,st941,rm2114,kmk1001,mjfg100\}@cam.ac.uk}
\keywords{under-represented, low-resource, speech recognition, foundation models, elastic weight consolidation}

\begin{document}

\maketitle

\footnotetext[1]{This paper reports on research supported by EPSRC Project EP/V006223/1 (Multimodal Video Search by Examples) and Cambridge University Press \& Assessment, a department of The Chancellor, Masters, and Scholars of the University of Cambridge.}

\begin{abstract}

Foundation ASR models often support many languages, e.g. 100 languages in Whisper. However, there has been limited work on integrating an additional, typically low-resource, language, while maintaining performance on the original language set. Fine-tuning, while simple, may degrade the accuracy of the original set. We compare three approaches that exploit adaptation parameters: soft language code tuning, train only the language code; soft prompt tuning, train prepended tokens; and LoRA where a small set of additional parameters are optimised. Elastic Weight Consolidation (EWC) offers an alternative compromise with the potential to maintain performance in specific target languages. Results show that direct fine-tuning yields the best performance for the new language but degrades existing language capabilities. EWC can address this issue for specific languages. If only adaptation parameters are used, the language capabilities are maintained but at the cost of performance in the new language.

\end{abstract}

\section{Introduction}

Recent years have seen impressive advancements in Automatic Speech Recognition (ASR) systems, particularly for languages with abundant linguistic resources, leading to high performance. However, the situation is different for ASR systems tailored to low-resource languages, the performance of which still falls short when compared to their high-resource counterparts.  Low-resource languages typically lack a strong online presence, linguistic expertise, sufficient speech and text data, and pronunciation lexicons. These characteristics pose challenges for developing effective ASR systems in such languages.

In the era of hybrid DNN-HMM models, extensive research on ASR for low-resource languages focused on building propriety lexicons for modelling the acoustic units attracts. \cite{qian2022automatic, lonergan2022cross} proposed a global lexicon, which captures dialect variant forms with relatively abstract representations, and a multi-dialect lexicon containing all dialect variants for Irish. 
Another approach was to develop graphemic lexicons where the ``pronunciation'' for a word is defined by the letters forming the word e.g. \cite{gales2015unicode}. End-to-end ASR models don't require a lexicon, so this doesn't affect performance. However, the lack of transcribed training data, as for the hybrid models, is a big problem for low-resources languages.



To handle the lack of training data, a common method is using self-supervised or semi-supervised training
~\cite{zhu22c_interspeech,synnaeve2020end, vesely2017semi}, where a model initially trained with limited transcribed data is used to generate transcriptions for unlabeled data which are then added to the training data to update the model. Data augmentation can also be used to tackle this problem where additional data is synthesised from existing data e.g.~\cite{xie23_interspeech,bartelds2023making,park19specaug,ragni14data}.

With the development and release of pre-trained multilingual end-to-end foundation ASR models~\cite{radford2023robust,zhang2023google,babu2021xls}, utilising a multilingual model is another option for low-resource languages. Zhang et al. proposed a smaller universal monolingual output layer shared across languages for high-quality and high-efficiency multilingual ASR~\cite{zhang2023uml}.
Chen et al. used hierarchical CTC to leverage language identity throughout the entire encoder-decoder network, aiming to improve ASR performance by correctly identifying languages~\cite{chen2023improving} . Qin et al. utilised multilingual and multilevel unit modeling to improve ASR performance on low-resource Tibetan languages~\cite{qin2022improving}.

The challenge in multilingual models lies in expanding them to new languages without compromising the performance of existing ones, particularly when the training data for new languages is limited or unavailable. 
In NLP, approaches like language-specific adapters and output heads, as seen in~\cite{he2021towards, houlsby2019parameter}, enable parameter-efficient fine-tuning but necessitate prior knowledge of the input language. Alternatively, continual lifelong learning, as proposed by \cite{parisi2019continual}, combines data from both existing and new languages for ongoing training. Demonstrating its effectiveness in ASR, \cite{li2021scaling,li2022massively} have expanded language coverage from 15 to 32 and to 66 languages, respectively.
Pham et al. proposed a weight factorization technique to factorize each weight matrix in the network into language dependent and independent factors~\cite{pham21efficient}. When combined with Elastic Weight Consolidation (EWC), this approach allows a multilingual ASR model to expand its learning from an initial 10 languages to 26 languages without catastrophic forgetting~\cite{pham23towards,Pham2023thesis}.


In this paper, we focus on adding a new language to an ASR foundation model, using Whisper as a case study. Firstly, the zero-shot ability of the foundation model is examined. Secondly, we compare efficient fine-tuning methods such as Low Rank Adaptation (LoRA) with standard fine-tuning. Moreover, we propose Soft Language Code Tuning (SLCT) to train a language-specific embedding vector for the new language. 
Inspired by techniques from NLP tasks, we implement Soft Prompt Tuning (SPT), introduced in~\cite{lester2021power} and utilised in a language assessment task~\cite{ma2023adapting_slate}, to effectively expand language coverage.
For fine-tuning, we analysed catastrophic forgetting and mitigated it with Elastic Weight Consolidation (EWC). Initially, we assess the zero-shot ability of Whisper on six languages not originally supported in the model, then focus on three languages to evaluate the performance of various approaches.

\section{Methods}
\label{sec:methods}
\subsection{Fine-tuning}
\label{sec:ft}
If the parameters of the ASR model are denoted as $\theta_{\text{ASR}}$, given a set of acoustic features, the goal of the ASR model is to produce the transcription $\hat Y = \{\hat y_1, \hat y_2, ..., \hat y_t\}$, which is determined by the equation
    $\hat Y = \mathop{\arg\max}_Y P(Y|X;\theta_{\text{ASR}})$.
Model fine-tuning (FT) is a standard approach for domain adaptation, where all model parameters are tuned using domain-specific data to minimise the training loss.
In this work, we adopt the fine-tuning method to integrate a new target language into  Whisper.
Benefiting from pre-training on an extensive set of labelled audio-transcription data, Whisper generally requires minimal additional fine-tuning to achieve an effective ASR model tailored for the target domain. 

\subsection{Efficient Fine-tuning with LoRA}
\label{sec:lora}
Low Rank Adaptation (LoRA) introduces an innovative and efficient fine-tuning strategy by leveraging low-rank approximation to adapt models to new data distributions~\cite{hu2021lora}.
We denote the parameters of the ASR model as $\theta_{\text{ASR}}=[\theta_{1}, \theta_{2}, ..., \theta_{L}]$, where each $\theta_{l}$ represents a weight matrix. Instead of directly updating, LoRA constrains the update of a weight matrix $\theta_{l} \in R^{d\times k}$ with a low-rank decomposition, such that $\theta_{l} + \Delta \theta_{l} = \theta_{l} + B_{l} \times A_{l}$, where $B_{l} \in R^{d\times r}$, $A_{l} \in R^{r\times k}$ and the rank $r \ll min(d, k)$.
This introduces new weight matrices $B$ and $A$ specific to the adapted domain, while preserving a set of original model parameters $\theta_{\text{ASR}}$ and ensuring the performance on all original tasks is unchanged.

\subsection{Soft Language Code Tuning}
\label{sec:slct}
\begin{figure}
    \centering
    \includegraphics[width=1\linewidth]{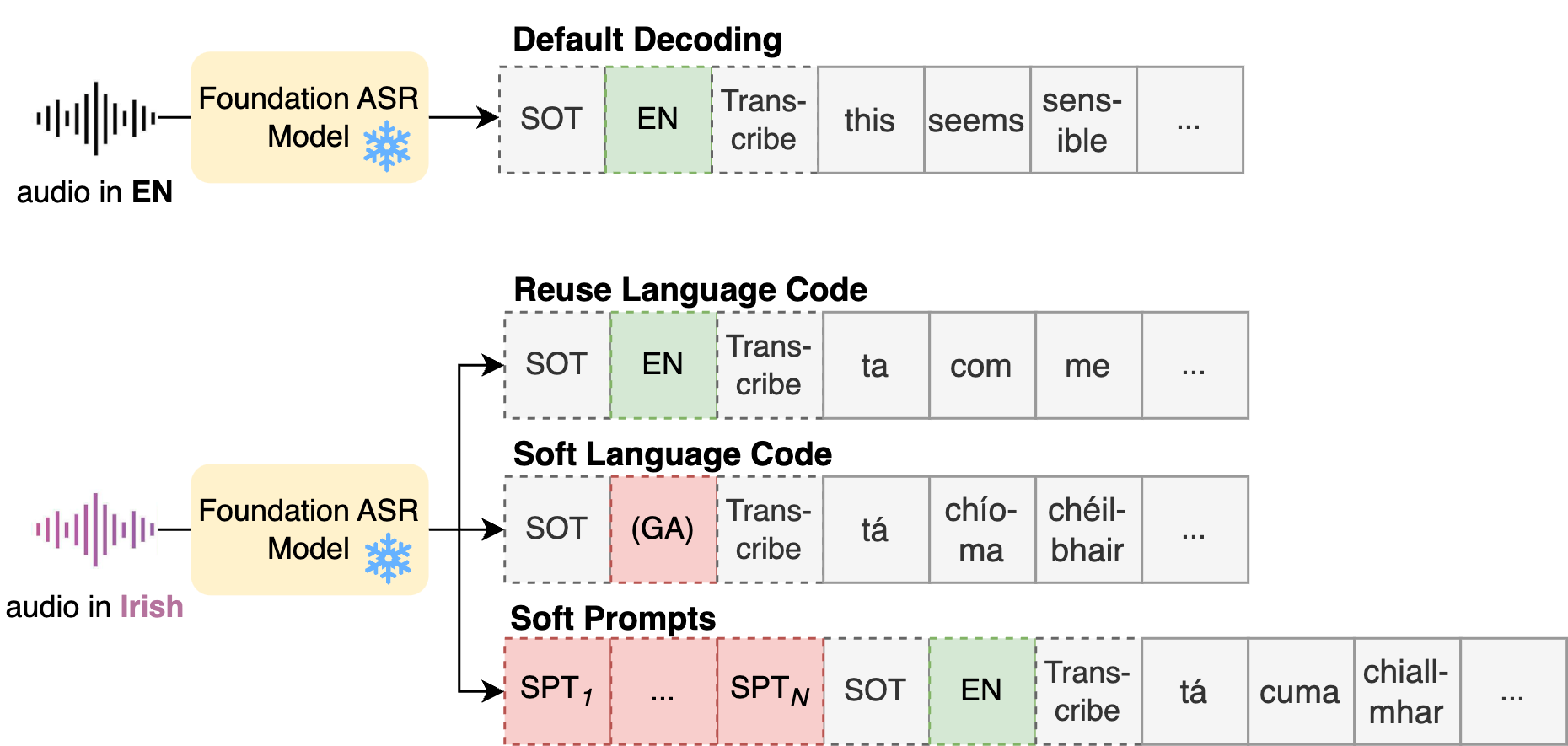}
    \caption{Illustration of the decoding process for existing and new languages using the default decoding and three methods: (a) reusing an existing language code, (b) using a trained soft language code, and (c) employing prepended soft prompts to capture new domain knowledge. Reference text for Irish audio: ``tá cuma chiallmhar air seo''.}
    \label{fig:soft-lang_soft-prompt}
\end{figure}

Whisper identifies each language by a unique token during its training phase~\cite{radford2023robust}. For an unseen language, however, this language code does not exist. To tackle this challenge, we introduce the concept of Soft Language Code Tuning (SLCT), depicted in Figure~\ref{fig:soft-lang_soft-prompt}. Rather than using the pre-trained embeddings for a specific language code, SLCT trains a new embedding vector for the target language code with the training set of the target language, while keeping other model parameters fixed. The soft language embeddings are exclusively employed for the target language, effectively incorporating a new language into the model while preserving the model's performance on prior tasks.



\subsection{Soft Prompt Tuning}
\label{sec:spt}
Prompting~\cite{liu2023pre, shin2020autoprompt} has become a popular method to adapt large language models (LLM)~\cite{kenton2019bert, raffel2020exploring, OpenAI2023GPT4TR}. Beyond the traditional discrete prompts, soft prompt tuning (SPT) employs continuous vectors that are inserted into the model's input and optimised via gradient descent on training data~\cite{lester2021power}. This approach enhances model performance and requires minimal human expertise.
\cite{ma2023adapting_slate} adopted soft prompt tuning for the ASR domain and demonstrated its effectiveness on the spoken language assessment task. Inspired by this, our work extends the concept of SPT to incorporate new languages into a foundation ASR model.
Specifically, $m$ token embeddings $V=\{v_0, ..., v_{m-1}\}$ were prepended to the decoder input embeddings. During training, only these embeddings $V$ are updated based on the loss function while keeping the original model parameters, $\theta_\text{ASR}$, unchanged. In this way, extra information is introduced for the model to condition on during the generation process, where 
$\hat Y = \mathop{\arg\max}_Y P(Y| X; \theta_{\text{ASR}}, \theta_V)$. An illustration of the modified model input to the ASR decoder during training and decoding is shown in Figure~\ref{fig:soft-lang_soft-prompt}.

\subsection{Forgetting and Elastic Weight Consolidation}
\label{sec:ewc}
LoRA, SLCT, and SPT all have the ability to maintain performance on prior domains or tasks as they do not alter the original model parameters. 
However, model fine-tuning carries the risk of catastrophic forgetting, a common challenge faced by large foundation models, which occurs when the model's proficiency in prior tasks declines as it undergoes continuous updates or fine-tuning for new tasks or domains~\cite{mccloskey1989catastrophic,french1999catastrophic,ratcliff1990connectionist,goodfellow2014empirical}. Various strategies to mitigate this issue comprises Pseudo-rehearsal~\cite{robins1995catastrophic,atkinson2021pseudo},  Memory Aware Synapses~\cite{aljundi2018memory} and Elastic Weight Consolidation (EWC)~\cite{kirkpatrick2017overcoming}.
In this study, we adopt EWC both as an analytic tool and as a regularisation approach to address the problem of catastrophic forgetting when adding a new language to the foundation ASR model.

EWC employs the observed Fisher information matrix $F$ to identify important parameters relative to the previous tasks during the pre-training phase. Fine-tuning is then modified to prevent large deviations in these important parameters. For a previous task $T_{A}$ (e.g. ASR for an existing language) and a new task $T_{B}$ (e.g. ASR for a new language), the loss function $L$ that we minimise in EWC is defined as follows:
\begin{equation}
\label{eq:ewc}
    L(\theta) = L_{B}(\theta) + \lambda\sum_{i}  F_{i}(\theta_{i} - \theta_{A,i}^{*})^2
\end{equation}
where $L_{B}(\theta)$ is the loss for task $T_{B}$ only, $\lambda$ sets the importance of the previous task $T_{A}$ relative to the new task $T_{B}$, for each parameter $i$. Here, $F_{i}$ represents the $i$-th diagonal element of the Fisher information matrix, $\theta_{i}$ is the $i$-th parameter of the model, and $\theta_{A,i}^{*}$  is the optimal value of the $i$-th parameter after training on the previous task $T_{A}$.

In addition to using EWC as a training constraint, we also utilise Fisher overlap~\cite{kirkpatrick2017overcoming}, a measure of the degree of overlap between the two tasks' Fisher matrices, as an analytic tool to examine potential forgetting of previous languages when integrating a new language into Whisper. The observations derived from this analysis can be tested through EWC fine-tuning.
Precisely, we computed the Fisher matrices associated with two languages, $F_{1}$ and $F_{2}$, and normalised these to each have unit trace, $\hat{F_{1}}$ and $\hat{F_{2}}$, then computed their Fr\'echet distance, a metric on the space of positive-semidefinite matrices:
\begin{equation}
\begin{aligned}
    d^{2}(\hat{F_{1}}, \hat{F_{2}}) &= \frac{1}{2} \times ||\hat{F_{1}}^{1/2} - \hat{F_{2}}^{1/2}||_{F}
\end{aligned}
\end{equation}
which is bounded between 0 and 1. We then define the overlap as $1-d^{2}$, where a value of zero indicates that the two tasks depend on non-overlapping sets of weights, while a value of one indicates that $F_{1} = \alpha F_{2} $ for some $\alpha > 0$.

\section{Experiments}
\subsection{Datasets}

Experiments were conducted on the FLEURS benchmark dataset~\cite{conneau2023fleurs}, which comprises 102 languages, of which 19 languages are not supported by the Whisper large-v3 model. Each language has about 12 hours of read speech data, with each audio within 30 seconds. Initial tests were performed on six languages randomly selected from the 19 not supported ones: Irish, Asturian, Sorani Kurdish, Cebuano, Kabuverdianu, and Kyrgyz. 
We then focused on the former three languages (Irish, Asturian, and Sorani Kurdish) for further investigation, selected for their diverse range of zero-shot ASR performance in terms of Character Error Rates (CER). The number of files and hours of the datasets of these three languages are presented in Table~\ref{tab:fleurs_trainset}. 
Additionally, our proposed approaches are also tested on languages in Whisper's supported list using the FLEURS dataset, including English (en), Spanish (es), Welsh (cy), and Occitan (oc). Both the original raw transcript and normalisation version with cases lowering and punctuation removing are provided in the dataset, the latter was used as an evaluation reference.
Three out-of-domain datasets are utilised to evaluate models' forgetting behavior, namely (1) Librispeech (LB), an audio read speech corpus~\cite{panayotov2015librispeech}, (2) Artie Bias Corpus (Artie), a subset of Common Voice Corpus~\cite{meyer2020artie}, and (3) TED-LIUM (Ted) test set which contains audio from TED conference videos~\cite{hernandez2018ted}.



\begin{table}[!htbp]
\center
\caption{FLEURS datasets used in training.} 
\label{tab:fleurs_trainset}
\begin{tabular}{l|cr|cr|cr}
\toprule 
 & \multicolumn{2}{c|}{Train} & \multicolumn{2}{c|}{Dev} & \multicolumn{2}{c}{Test}\\
Language & \# & hrs & \# & hrs & \# & hrs\\
\midrule
Irish & 2,845 & 12.12  & 369 & 1.49 & 842 & 3.46 \\
Asturian  & 2,511 & 7.54 & 398 & 0.99 & 946 & 2.44 \\
Sorani Kurd & 3,040 & 10.46 & 386 & 1.23 & 922 & 2.99 \\
\bottomrule
\end{tabular}
\end{table}



\label{sec:setup}
Whisper has various model sizes. We use the latest version large-v3 as the foundation model because it shows improved performance over a wide range of languages compared with other models\footnote{https://github.com/openai/whisper/discussions/1762}. Fine-tuning (FT) employs an initial learning rate of 1e-5 and decays linearly, while SPT and SLCT use initial rates of 1e-4 and 1e-1, respectively. SPT results are presented with 20 soft prompts.
LoRA was applied to attention layers with a rank of 8, with varying rank sizes showing negligible differences in results.
EWC experiments were conducted with $\lambda$ values (in Equation~\ref{eq:ewc}) selected from the range of [1e-0, 1e-1, 1e-2, 1e-3, 1e-4, 1e-5]. The optimal $\lambda$ was determined on the development set.


\subsection{Zero-shot ASR and Speech Translation}
\label{sec:zero-shot}
The Whisper large-v3 model has been pre-trained on 100 languages for ASR, Speech Translation (ST) and Language Identification (LID) tasks; exposed to extensive acoustic and language data, it prompts an intriguing question regarding its capability for zero-shot ASR or ST for an unseen language. To investigate this, we performed zero-shot ASR and ST on six languages not supported by Whisper with the FLEURS test sets on the large-v3 model. Results, presented in Table~\ref{tab:zero}, reveal a diverse range of CER, yet consistently poor Word Error Rates (WER) across all languages, underscoring the challenges of zero-shot ASR for unseen languages. 

For ST evaluation, we use BLEU and COMET scores, specifically using the wmt22-comet-da~\cite{rei2022comet} model to calculate the COMET score. Although wmt22-comet-da was trained on numerous languages, it lacks Asturian, Cebuano, Kabuverdianu, and Sorani Kurdish, potentially affecting COMET scores for these languages.
Whisper achieved notably high BLEU scores, especially for Asturian and Kabuverdianu,
despite not being trained on these languages, demonstrating its capability for zero-shot speech translation on unseen languages. 
Since ASR appears to be more challenging for under-represented languages, we focus on the ASR task for the rest of the paper.

\begin{table}[!t]
    \caption{Whisper large-v3 model zero-shot performance for under-represented languages on the ASR and ST tasks.}
    \label{tab:zero}
    \centering
    \begin{tabular}{l|cc|cc}
    \toprule
    \multirow{2}{*}{Language} & \multicolumn{2}{c|}{ASR} & \multicolumn{2}{c}{ST} \\
     & CER & WER & BLEU & COMET \\
    \midrule
    Asturian & 13.5 & 47.5 & 27.6 & 73.1 \\
    Cebuano & 47.7 & 72.9 & 8.8 & 55.1\\
    Kabuverdianu & 35.2 & 87.8 & 32.3 & 73.7 \\
    Kyrgyz & 50.0 & 87.4 & 4.7 & 56.5 \\
    Sorani Kurdish & 38.6 & 110.8 & 2.2 & 45.6\\
    Irish   & 78.8 & 110.4 & 1.7 & 43.1 \\
    \bottomrule
    \end{tabular}
\end{table}

%


\subsection{Language code selection}
\label{sec:exp_irish}
\begin{table}[!b]
    \centering
    \caption{ASR performance on FLEURS Irish test set with the baseline and the fine-tuned Whisper large-v3 model using different language code (en: English, cy: Welsh, oc: Occitan, es: Spanish, ga: Irish).}
    \label{tab:whisper_asr_irish}
    \begin{tabular}{c|cc|cc}
    \toprule
    Language & \multicolumn{2}{c|}{Baseline} & \multicolumn{2}{c}{FT} \\
    Code & CER  & WER &  CER & WER \\
    \midrule
    en  & 78.8 & 110.4 & 22.3 & \bf43.7\\
    cy & 62.3 & 108.0 & \bf22.2 & 44.0 \\
    oc & \textbf{45.6} & \textbf{90.1} & 22.4 & 44.2 \\
    es & 79.2 & 107.3 & 22.6 & 44.7 \\
    ga & - & - & 22.8 & 44.5 \\
    \midrule
    SLCT & \textbf{41.5} & \textbf{81.4} & - & -\\
    \bottomrule
    \end{tabular}
\end{table}

\noindent
Whisper leverages a language code for decoding a supported language. However, there is no corresponding language code when a low-resource language is not supported. One straightforward approach to address this limitation is to find a related source language from the supported list and utilise its language code for inference on the target low-resource language.  
Taking Irish (`ga') as an illustration, we selected four language codes associated with languages exhibiting the highest character and token overlaps with Irish.
In addition, we applied Soft Language Code Tuning (SLCT), to train an Irish-specific embedding vector specific.
Table~\ref{tab:whisper_asr_irish} presents results on the FLEURS Irish test set. 
The choice of language code can be seen to impact ASR performance, showing an absolute difference of 33.2\% in CER and 20.3\% in WER. Similar findings were noted on Asturian and Sorani Kurdish languages. SLTC gives the best performance, setting an upper bound for language code selection.



In contrast to previous findings, fine-tuning proves to be robust to variations in language code, delivering large performance gains regardless of the chosen language code, as presented in Table~\ref{tab:whisper_asr_irish}. With the `en' language code, we observe an improvement of 56\% and 66\% in CER and WER, respectively, on the test set. 
Interestingly, when introducing a new language code (`ga') for training, we observe slightly worse performance. This is likely due to the fact that the `ga' code is new to the model, suggesting a potential need for more training data to effectively integrate a new language code.

\subsection{Efficient fine-tuning}



Table~\ref{tab:whisper_asr_other_tuning} presents a comparison of different fine-tuning approaches across three target languages: Irish, Asturian, and Sorani Kurdish. Baseline performance is obtained using the same Whisper large-v3 model for all three languages, while other models (FT, SLCT, SPT, and LoRA) are specifically trained for each language using the respective FLEURS training sets.
For Irish, FT achieves the best performance improvement, with a 56.5\% absolute reduction in CER and a 66.7\% reduction in WER compared to other efficient fine-tuning approaches. This trend is consistent with the other two languages, which is unsurprising as the model is tailored to a specific domain for each target language, adjusting all parameters from the original domain.
SLCT results in performance gains of 26.3\%, 37.3\%, and 14.8\% on Irish, Asturian, and Sorani Kurdish, respectively. Despite the relatively low performance, it's worth noting that only 0.00008\% of the model's 1550M parameters are tuned, showcasing noticeable achievement given the tuned parameter size.
SPT, which tunes 0.001\% of the model parameters, further enhances the performance by 32.2\%, 59.2\%, and 44.6\% relative on Irish, Asturian, and Sorani Kurdish, respectively.
LoRA, tuning 0.13\% of model parameters, demonstrates improved performance proportionate to the increased size of trained parameters.

\begin{table}[!htbp]
    \caption{ASR performance on FLEURS Irish, Asturian, and Sorani Kurdish test sets with different parameter-tuning approaches.}
    \label{tab:whisper_asr_other_tuning}
    \centering
    \begin{tabular}{@{ }l@{ }|cc|cc|cc}
    \toprule
    \multirow{2}*{Model} & \multicolumn{2}{c|}{Irish} & \multicolumn{2}{c|}{Asturian}  & \multicolumn{2}{c}{Sorani Kurdish} \\
     & CER   & WER  &  CER  & WER &  CER  & WER \\
    \midrule   
    Baseline &  78.8 & 110.4 & 13.5 & 47.5 & 38.6 & 110.8 \\
    FT & 22.3 & 43.7 & 3.7 & 11.4 & 7.3 & 32.9 \\
    \midrule
    SLCT & 41.5 & 81.4 & 7.4 & 29.8 & 35.8 & 95.2\\
    SPT & 38.1 & 74.9 & 5.1 & 19.4 & 13.6 & 61.4 \\
    LoRA  & 28.3 & 54.8 & 4.1 & 14.0  & 8.7 & 41.3 \\
    \bottomrule
    \end{tabular}
\end{table}

\subsection{Forgetting and EWC}
\begin{figure}
    \centering
    \includegraphics[width=0.92\linewidth]{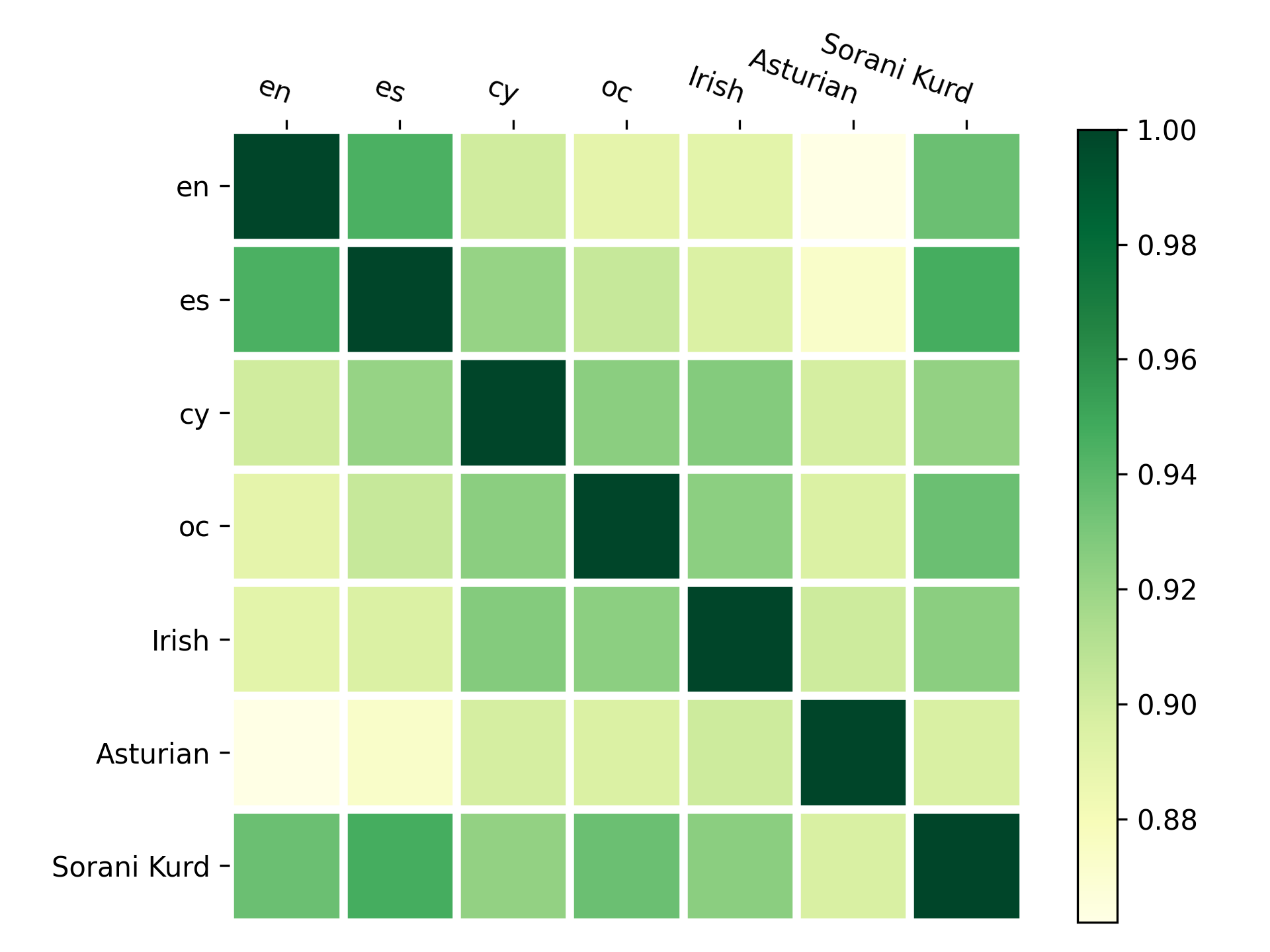}
    \caption{Fisher overlap between Fisher parameters extracted for different languages or datasets.}
    \label{fig:fisher}
\end{figure}

As discussed in Section~\ref{sec:ewc}, fine-tuning a model on a specific target language can lead to a performance decline in other languages. 
To demonstrate this effect, we analysed the Fisher overlap among the test sets for various languages, as depicted in Figure~\ref{fig:fisher}.
Notably, Irish exhibits a higher Fisher overlap with Welsh (cy) and Occitan (oc) than with English (en) and Spanish (es), indicating a greater parameter overlap between Irish and Welsh (or Occitan). This implies that fine-tuning the model on Irish could adversely affect its performance on Welsh, a finding confirmed by the results in Table~\ref{tab:whisper_asr_irish_other} where the WER for Welsh (cy) decreases from 27.5\% to 96.8\%, the biggest degradation compared to other languages. 
When training is constrained with EWC Fisher parameters extracted from the FLEURS English training set, we note a decline in performance on the Irish test set but improvements across all other languages compared to standard fine-tuning. It even manages to preserve the performance on out-of-domain English test sets, halving the WER degradation compared to standard fine-tuning. However, maintaining performance on cy proves more challenging than for en or es due to the higher Fisher overlap between Irish and cy.
By constraining the fine-tuning with EWC Fisher parameters extracted with the FLEURS cy training set with a $\lambda$ value set to 1, we observe better performance preservation on cy but a small decline in performance on Irish. This highlights the difficulty in enhancing performance on two tasks with high Fisher overlap.

\begin{table}[!htbp]
\caption{Evaluation of WER on in-domain and out-of-domain test sets using Whisper large-v3 and its fine-tuned version. The model is fine-tuned with FLEURS Irish training set. Optimal $\lambda$ for EWC (en): 1e-1; optimal $\lambda$ for EWC (cy): 1.0.}
    \label{tab:whisper_asr_irish_other}
    \centering
    \begin{tabular}{@{}l@{ }|C{5.5mm}|C{3.0mm}C{3.0mm}C{3.7mm}C{4.0mm}|C{3.0mm}C{3.7mm}c@{}}
    \toprule 
    Model & \multicolumn{5}{c|}{FLEURS} & \multicolumn{3}{c}{Out-of-domain} \\
     & Irish  & en  & es & cy & oc & LB & Artie & Ted\\
    \midrule
    Baseline & 110.4 & 4.1 & 2.8 & 27.5 & 64.2 & 3.7 & 5.6 & 4.0 \\
    FT & 43.7 & 6.1 & 4.1 & 96.8 & 79.8 & 8.8 & 13.0 & 4.2 \\
    \hspace{0.6mm} + EWC (en) & 46.7 & 4.6 & 2.8 & 78.7 & 71.5 & 5.4 & 9.0 & 4.0 \\ 
    \hspace{0.6mm} + EWC (cy) & 47.0 & 4.5 & 3.0 & 66.1 & 71.3 & 5.4 & 8.8 & 4.0 \\ 
    \bottomrule
    \end{tabular}
\end{table}

\section{Conclusion}
In this paper, we examine Whisper's zero-shot ASR and ST capabilities on unseen languages and then focus on adding new languages into Whisper for the ASR task. In addition to the implementation of LoRA, we introduce SLCT and SPT for efficient fine-tuning. While standard fine-tuning achieves the best performance compared to various efficient fine-tuning approaches, it faces the challenge of catastrophic forgetting. To address this, we adopt Fisher overlap as an analytic tool for assessing potential forgetting on different languages. Employing EWC of Fisher parameters, we can constrain forgetting and preserve performance on existing languages. Our experiments reveal that it's hard to learn a new task and not forget a previous task if two tasks have a high Fisher overlap. In future work, we will extend the investigation into speech translation and explore integrating a new language into foundation models for the translation task while ensuring that performance on previous languages or tasks is preserved.



\bibliographystyle{IEEEtran}
\bibliography{mybib}

\end{document}